\documentstyle[12pt]{article}
\renewcommand{\thefootnote}{\fnsymbol{footnote}}

\newlength{\extraspace}
\newlength{\extraspaces}
\newcommand{\be}{\begin{equation}
\addtolength{\abovedisplayskip}{\extraspaces}
\addtolength{\belowdisplayskip}{\extraspaces}
\addtolength{\abovedisplayshortskip}{\extraspace}
\addtolength{\belowdisplayshortskip}{\extraspace}}
\newcommand{\ee}{\end{equation}}
\newcommand{\ba}{\begin{eqnarray}
\addtolength{\abovedisplayskip}{\extraspaces}
\addtolength{\belowdisplayskip}{\extraspaces}
\addtolength{\abovedisplayshortskip}{\extraspace}
\addtolength{\belowdisplayshortskip}{\extraspace}}
\newcommand{\ea}{\end{eqnarray}}
\newcommand{\nonu}{\nonumber \\[.5mm]}
\newcommand{\A}{&\!\!\!}

\begin{document}
\addtolength{\baselineskip}{3.0mm}

\begin{center}
{\large{\bf{Generalized Lagrangian of 
            $N = 1$ supergravity 
            
            and 
            
            its canonical constraints 
            with the real Ashtekar variable}}} 
\\[12mm]
{\large Motomu Tsuda}
\footnote{e-mail: mtsuda@krishna.th.phy.saitama-u.ac.jp} 
\\[7mm]
Physics Department, Saitama University \\[2mm]
Urawa, Saitama 338, Japan \\[12mm]
{\bf Abstract}\\[7mm]
{\parbox{13cm}{\hspace{5mm} 
We generalize the Lagrangian of $N = 1$ supergravity 
(SUGRA) by using an arbitrary parameter $\xi$, 
which corresponds to the inverse of Barbero's parameter 
$\beta$. This generalized Lagrangian involves 
the chiral one as a special case of the value $\xi = \pm i$. 
We show that the generalized Lagrangian gives 
the canonical formulation of $N = 1$ SUGRA 
with the real Ashtekar variable 
after the 3+1 decomposition of spacetime. 
This canonical formulation is also derived 
from those of the usual $N = 1$ SUGRA 
by performing Barbero's type canonical transformation 
with an arbitrary parameter $\beta \ (= \xi^{-1})$. 
We give some comments on the canonical formulation 
of the theory.}} 
\end{center}
\vfill

\newpage

\renewcommand{\thefootnote}{\arabic{footnote}}
\setcounter{footnote}{0}

A very simple, polynomial form of Hamiltonian 
constraint in canonical formulation of general relativity 
is obtained by using the complex Ashtekar's connection 
variable \cite{AA}. However, it is difficult to 
deal with the reality condition especially at quantum 
level, which must be imposed in order to select 
the physical, Lorentzian theory \cite{AA2}. 
A possible way to solve this problem of the reality 
condition was proposed by Barbero \cite{Bar} 
using the real-valued Ashtekar variable, 
although one must discard the polynomiality of 
the Hamiltonian constraint in the Lorentzian sector. 

The advantage of the formulation with the real Ashtekar 
variable in pure gravity is that it provides 
a mathematically rigorous kinematical framework 
in the context of diffeomorphism invariant quantization 
with the Gauss and vector constraints being 
satisfied \cite{ALMMT}. Furthermore, Thiemann has 
recently succeeded in constructing a quantum 
Hamiltonian constraint operator which is mathematically 
well-defined in the Lorentzian sector \cite{TT1}. 

Canonical formulation of general relativity 
with the real Ashtekar variable has been made 
starting from the generalized Einstein-Cartan (EC) 
action \cite{Ho}.
\footnote{\ In Ref.\cite{Ho} this generalized action 
is called the generalized Hilbert-Palatini action, 
which corresponds to the generalization of the tetrad 
form of the Palatini action.} 
In this paper we extend the action to include 
spinor matter fields, and then derive the canonical 
formulation of $N = 1$ supergravity (SUGRA) 
with the real Ashtekar variable. 

We begin with briefly reviewing the generalization 
of the EC Lagrangian \cite{Ho}. We denote the tetrad 
field as $e^i_{\mu}$, from which the metric field 
$g_{\mu \nu}$ is constructed via 
$g_{\mu \nu} = \eta_{ij} e^i_{\mu} e^j_{\nu}$.
\footnote{\ Greek letters $\mu, \nu, \cdots$ are 
spacetime indices, and Latin letters {\it i, j,} $\cdots$ 
are local Lorentz indices. 
We denote the Minkowski metric 
by $\eta_{ij} = {\rm diag} (-1, +1, +1, +1)$. 
The totally antisymmetric tensor $\epsilon_{ijkl}$ 
is normalized as $\epsilon_{0123} = +1$. 
We define $\epsilon_{\mu \nu \rho \sigma}$ 
and $\epsilon^{\mu \nu \rho \sigma}$ as tensor 
densities which take values of $+1$ or $-1$.} 
The Lorentz connection $A_{ij \mu}$ is treated as 
independent variable in the EC Lagrangian. 
Then the generalized EC Lagrangian density, 
which derives Barbero's results using the real 
Ashtekar variable, takes the form 
\be
{\cal L}_G = {e \over 2} \ e_i^{\mu} e_j^{\nu} 
             \left( R{^{ij}}_{\mu \nu} - {\xi \over 2} 
             \epsilon{^{ij}}_{kl} R{^{kl}}_{\mu \nu} \right), 
\label{gHP}
\ee
where $e = {\rm det}(e^i_{\mu})$ and $R{^{ij}}_{\mu \nu}$ 
is the curvature tensor with respect to the Lorentz 
connection $A_{ij \mu}$. Here a complex parameter $\xi$ 
is introduced in (\ref{gHP}) in order to cover various 
types of the canonical formulation of general relativity:
\footnote{\ The parameter $\xi$ is same as the parameter 
$\alpha$ of \cite{Ho} and corresponds to the inverse 
of Barbero's parameter $\beta$ as stated in \cite{Ho}.} 
Indeed, for $\xi = 0$, Eq.(\ref{gHP}) is simply the EC 
Lagrangian which leads to the ADM canonical formulation. 
For $\xi = + i \ (- i)$, only the self-dual 
(antiself-dual) part of the curvature contributes 
to the Lagrangian density (\ref{gHP}).
\footnote{\ We denote the self-dual and antiself-dual 
part of a antisymmetric tensor $F_{ij}$ as 
$F^{(\pm)}_{ij}$ which satisfies 
$(1/2){\epsilon_{ij}} \! ^{kl} F^{(\pm)}_{kl} 
= \pm i F^{(\pm)}_{ij}$.} 
In this case the complex (anti)self-dual connection 
$A^{(\pm)}_{ij \mu}$ is regarded as an independent 
variable and then Eq.(\ref{gHP}) leads to Ashtekar's 
canonical formulation after the 3+1 decomposition 
of spacetime. On the other hand, the canonical formulation 
using the real Ashtekar variable is derived 
by putting $\xi$ to be real.
\footnote{\ Immirzi pointed out \cite{Imm} that 
the Barbero's parameter $\beta \ (= \xi^{-1})$ appears 
as a free (real) parameter in the quantum spectrum 
of such geometrical quantities as length, 
area and volume. Therefore the parameter $\beta$ 
is also known as the Immirzi parameter.} 

Such a generalization of the EC Lagrangian as given by 
(\ref{gHP}) does not affect the field equation 
for the tetrad in the second-order formalism \cite{Ho}. 
In order to see this, it is convenient to introduce 
the variable 
\be
B_{ij \mu} := {1 \over 2} \left( A_{ij \mu} 
- {\xi \over 2} \epsilon{_{ij}}^{kl} A_{kl \mu} \right), 
\label{Bdef}
\ee
which reduces to $A^{(\pm)}_{ij \mu}$ 
for $\xi = \pm i$. Note that (\ref{Bdef}) can be 
solved with respect to $A_{ij \mu}$ 
unless $\xi = \pm i$. Since the variation of 
(\ref{gHP}) with respect to $A_{ij \mu}$ 
can be written as 
\be
\delta {\cal L}_G 
= - 2 \ D_{\mu} (e \ e_i^{\mu} \ e_j^{\nu}) 
\ \delta B{^{ij}}_{\nu}, 
\ee
we get the field equation for $A_{ij \mu}$,
\footnote{\ The antisymmetrization of a tensor with 
respect to $i$ and $j$ is denoted by 
$A_{[i \dots j]} := (1/2)(A_{i \dots j} 
- A_{j \dots i})$.} 
\be
D_{\mu} (e \ e_{[i}^{\mu} \ e_{j]}^{\nu}) = 0 
\label{A-eq}
\ee
with $D_{\mu}$ being the covariant derivative 
with respect to local Lorentz indices. 
The equation (\ref{A-eq}) is the same as that 
obtained from the EC Lagrangian, and can be solved 
to show that the $A_{ij \mu}$ is given by 
the Ricci rotation coefficient $A_{ij \mu}(e)$. 
Thus, in the second-order formalism, 
the second term in (\ref{gHP}) vanishes because of 
the Bianchi identity. This situation is just the same 
as in the case $\xi = \pm i$ \cite{JS}, 
and therefore (\ref{gHP}) is reduced to the ordinary 
Hilbert-Einstein Lagrangian of the tetrad form. 

Let us now try to introduce a (Majorana) 
Rarita-Schwinger field in the manner consistent 
with the above generalization of the EC Lagrangian. 
For this purpose, following the construction of 
the chiral Lagrangian of matter fields \cite{TSX}, 
we add a total divergence term with an arbitrary 
complex parameter $\eta$ to the ordinary Lagrangian 
of a (Majorana) Rarita-Schwinger field in flat space, 
and take the flat-space Lagrangian as 
\be
L_{RS} = L_{RS}({\rm ordinary}) 
         + {i \over 4} \ \eta \ \partial_{\mu} 
         (\epsilon^{\mu \nu \rho \sigma} 
         \overline \psi_{\nu} \gamma_{\rho} 
         \psi_{\sigma}) 
       = \epsilon^{\mu \nu \rho \sigma} \ 
         \overline \psi_{\mu} \gamma_5 \gamma_{\rho} 
         {{1 - i \eta \gamma_5} \over 2} 
         \partial_{\sigma} \psi_{\nu}. 
\label{fRS}
\ee
Then we apply the minimal prescription for (\ref{fRS}) 
replacing the ordinary derivative by the covariant 
derivative
\footnote{In our convention the Lorentz generator 
$S_{ij} = (i/4)[\gamma_i, \gamma_j]$ and 
$\{ \gamma_i, \gamma_j \} = -2 \eta_{ij}$.}
\be
D_{\mu} = \partial_{\mu} + {i \over 2} A_{ij \mu} S^{ij}, 
\ee
and define the generalized Lagrangian density 
of a (Majorana) Rarita-Schwinger field in curved space by 
\be
{\cal L}_{RS} = \epsilon^{\mu \nu \rho \sigma} 
                \overline \psi_{\mu} \gamma_5 \gamma_{\rho} 
                {{1 - i \eta \gamma_5} \over 2} 
                D_{\sigma} \psi_{\nu}. 
\label{gRS}
\ee
Notice that the right-hand side of (\ref{gRS}) 
agrees with the chiral Lagrangian density of 
a (Majorana) Rarita-Schwinger field when 
$\eta = \pm i$. 

We define the Lagrangian density ${\cal L}$ 
by the sum of (\ref{gHP}) and (\ref{gRS}), 
\be
{\cal L} := {\cal L}_G + {\cal L}_{RS}, 
\ee
and we require the ${\cal L}$ to reduce to 
the Lagrangian density of the usual $N = 1$ SUGRA 
in the second-order formalism. Varying the ${\cal L}$ 
with respect to $B_{ij \mu}$, we obtain 
\be
D_{\mu} (e \ e_{[i}^{\mu} \ e_{j]}^{\nu}) 
= {{1 + \xi \ \eta} \over {1 + \xi^2}} \ X{_{ij}}^{\nu} 
  + {{\xi - \eta} \over 2(1 + \xi^2)} 
  \ \epsilon{_{ij}}^{kl} X{_{kl}}^{\nu}, 
\label{ASG-eq}
\ee
where $X{_{ij}}^{\mu}$ is a tensor density defined by 
\ba
X{_{ij}}^{\mu} := \A \A {1 \over {1 + \eta^2}} 
                  \left( 
                  {{\delta {\cal L}_{RS}} \over 
                  {\delta A{^{ij}}_{\mu}}} 
                  + {\eta \over 2} \epsilon{_{ij}}^{kl} 
                  {{\delta {\cal L}_{RS}} \over 
                  {\delta A{^{kl}}_{\mu}}} 
                  \right) \nonu
                = \A \A {i \over 4} 
                  \epsilon^{\mu \nu \rho \sigma} \ 
                  \overline \psi_{\nu} \gamma_5 \gamma_{\rho} 
                  S_{ij} \psi_{\sigma}. 
\ea
If we substitute the solution of (\ref{ASG-eq}) with 
respect to $A_{ij \mu}$ into ${\cal L}$, then its torsion 
part gives four-fermion contact terms, 
which coincide with the contact terms of the usual 
$N = 1$ SUGRA if we choose $\eta = \xi$: In fact, 
we obtain
\footnote{The divergence term of (\ref{L-2nd}) is just 
the Chern-Simons type boundary term, which generates 
the chiral SUGRA `on shell' for $\xi = \pm i$ 
\cite{Mi1,Mi2}.}
\be
{\cal L}({\rm second \ order}) 
= {\cal L}_{N = 1 \ {\rm usual \ SUGRA}} ({\rm second \ order}) 
  + {i \over 4} \ \xi \ \partial_{\mu} 
  (\epsilon^{\mu \nu \rho \sigma} 
  \overline \psi_{\nu} \gamma_{\rho} \psi_{\sigma}) 
\label{L-2nd}
\ee
by means of the (Fierz) identity 
$\epsilon^{\mu \nu \rho \sigma} (\overline \psi_{\mu} 
\gamma_i \psi_{\nu}) \gamma^i \psi_{\rho} \equiv 0$. 
On the other hand, if $\eta \not= \xi$, 
these parameters survive in the contact terms 
as is seen from (\ref{ASG-eq}). 
Thus we shall assume that $\eta = \xi$ henceforth. 

The generalized Lagrangian density of $N = 1$ SUGRA 
in first-order form is now given by 
\be
{\cal L} = {e \over 2} e_i^{\mu} e_j^{\nu} 
           \left( R{^{ij}}_{\mu \nu} - {\xi \over 2} 
           \epsilon{^{ij}}_{kl} R{^{kl}}_{\mu \nu} \right) 
           + \epsilon^{\mu \nu \rho \sigma} \ 
           \overline \psi_{\mu} \gamma_5 \gamma_{\rho} 
           {{1 - i \xi \gamma_5} \over 2} 
           D_{\sigma} \psi_{\nu}, 
\label{gSG}
\ee
which is reduced to the Lagrangian density 
of $N = 1$ chiral SUGRA for $\xi = \pm i$. 
In case of the non-chiral theory with 
$\xi \not= \pm i$, the Lagrangian density of 
(\ref{gSG}) is invariant under the following 
first-order (i.e. `off-shell') SUSY transformations 
generated by a Majorana spinor parameter $\alpha$; 
namely, 
\ba
\A \A \delta \psi_{\mu} = D_{\mu} \alpha, 
\label{SUSY-psi} \\
\A \A \delta e^i_{\mu} = {i \over 2} \overline \alpha 
                         \ \gamma^i \psi_{\mu}, 
\label{SUSY-e} \\
\A \A \delta B_{ij \mu} = {1 \over 2}(C_{\mu ij} 
                          - e_{\mu [i} 
                          C{^m}_{mj]}), 
\label{SUSY-B}
\ea
where we define $C^{\lambda \mu \nu}$ as 
\be
C^{\lambda \mu \nu} := e^{-1} \epsilon^{\mu \nu \rho \sigma} 
                       \overline \alpha \ \gamma_5 
                       \gamma^{\lambda} 
                       {{1 - i \xi \gamma_5} \over 2} 
                       D_{\rho} \psi_{\sigma}. 
\ee
The transformations of (\ref{SUSY-psi}) and (\ref{SUSY-e}) 
are the same as those of the usual $N = 1$ SUGRA, 
whereas Eq.(\ref{SUSY-B}) differs from the usual one, 
since $C^{\lambda \mu \nu}$ depends 
on the parameter $\xi$. The form of (\ref{SUSY-B}), 
however, is easily read from the usual one 
if we note the relation 
\be
{{1 - i \xi \gamma_5} \over 2} A_{ij \mu} S^{ij} 
= B_{ij \mu} S^{ij} 
\ee
in the covariant derivative $D_{\sigma} \psi_{\nu}$ 
of (\ref{gSG}). In case of the chiral theory, 
however, the situation is slightly different: 
For example, when $\xi = +i$, Eq.(\ref{SUSY-B}) 
becomes the transformation of $A^{(+)}_{ij \mu}$, 
i.e. $\delta B_{ij \mu} \mid_{\xi = +i} 
= \delta A^{(+)}_{ij \mu}$, while $A^{(-)}_{ij \mu}$ 
which appears in (\ref{SUSY-psi}) is not an independent 
variable but a quantity given by $e^i_{\mu}$ 
and $\psi_{\mu}$ \cite{Jac,TS}. 

The generalized Lagrangian (\ref{gSG}) allows us to 
construct a canonical formulation of $N = 1$ SUGRA 
in terms of the real Ashtekar variable. 
Let us derive this by means of the Legendre transform 
of (\ref{gSG}) using the (3+1) decomposition of spacetime. 
For this purpose we assume that the topology of spacetime 
$M$ is $\Sigma \times R$ for some three-manifold $\Sigma$ 
so that a time coordinate function $t$ is defined on $M$. 
Then the time component of the tetrad can be defined as 
\footnote{Latin letters $a, b, \cdots$ are 
the spatial part of the spacetime indices 
$\mu, \nu, \cdots$, and capital letters 
{\it I, J,} $\cdots$ denote the spatial part 
of the local Lorentz indices {\it i, j,} $\cdots$.}
\be
e^i_t = N n^i + N^a e^i_a. 
\ee
Here $n^i$ is the timelike unit vector orthogonal to 
$e_{ia}$, i.e. $n^i e_{ia} = 0$ and $n^i n_i = - 1$, 
while $N$ and $N^a$ denote the lapse function 
and the shift vector, respectively. 
Furthermore, we give a restriction on the tetrad with 
the choice $n_i = (- 1, 0, 0, 0)$ in order to simplify 
the Legendre transform of (\ref{gSG}). 
Once this choice is made, $e_{Ia}$ becomes tangent to 
the constant $t$ surfaces $\Sigma$ and $e_{0a} = 0$. 
Therefore we change the notation $e_{Ia}$ 
to $E_{Ia}$ below. We also take the spatial restriction 
of the totally antisymmetric tensor 
$\epsilon^{\mu \nu \rho \sigma}$ 
as $\epsilon^{abc} := \epsilon{_t}^{abc}$, 
while $\epsilon^{IJK} := \epsilon{_0}^{IJK}$. 

Under the above gauge condition of the tetrad, 
the (3+1) decomposition of (\ref{gSG}) yields 
\ba
{\cal L} = \A \A 
           \epsilon^{abc} \epsilon_{IJK} 
           E^I_a (E^J_b \hat R{^{0K}}_{tc} 
           - N^d E^J_d \hat R{^{0K}}_{bc} 
           + {N \over 2} \hat R{^{JK}}_{bc}) \nonu
           \A \A 
           - \ \epsilon^{abc} (\overline \psi_b 
           \gamma_5 \gamma_c \hat D_t \psi_a 
           - \overline \psi_a 
           \gamma_5 \gamma_t \hat D_b \psi_c \nonu
           \A \A 
           + \ \overline \psi_t \gamma_5 \gamma_a 
           \hat D_b \psi_c 
           - \overline \psi_b \gamma_5 \gamma_c 
           \hat D_a \psi_t) 
\label{3+1dec}
\ea
with $\gamma_t = e^i_t \gamma_i = N \gamma_0 
+ N^a \gamma_a$ and $\gamma_a = E^I_a \gamma_I$. 
In (\ref{3+1dec}), $\hat D_{\mu}$ is defined by 
\be
\hat D_{\mu} := {{1 - i \xi \gamma_5} \over 2} D_{\mu}, 
\label{Dhat}
\ee
and also the quantity, $\hat R{^{ij}}_{\mu \nu} 
:= (1/2) \{ R{^{ij}}_{\mu \nu} - (\xi/2) 
\epsilon{^{ij}}_{kl} R{^{kl}}_{\mu \nu} \}$, 
is decomposed as 
\ba
\hat R{^{0K}}_{tc} = \A \A \partial_{[t} A{^{0K}}_{c]} 
      + A{^0}_{L [t} A{^{LK}}_{c]} 
      + {\xi \over 2} \ \epsilon^{IJK} 
      (\partial_{[t} A_{IJc]} \nonu
      \A \A 
      + \ A_{I0 [t} A{^0}_{Jc]} 
      + A_{IM [t} A{^M}_{Jc]}), 
      \label{R-dec1} \\
\hat R{^{0K}}_{bc} = \A \A \partial_{[b} A{^{0K}}_{c]} 
      + A{^0}_{L [b} A{^{LK}}_{c]} 
      + {\xi \over 2} \ \epsilon^{IJK} 
      (\partial_{[b} A_{IJc]} \nonu
      \A \A 
      + \ A_{I0 [b} A{^0}_{Jc]} 
      + A_{IM [b} A{^M}_{Jc]}), \\
\hat R{^{JK}}_{bc} = \A \A \partial_{[b} A{^{JK}}_{c]} 
      + A{^J}_{0 [b} A{^{0K}}_{c]} 
      + A{^J}_{I [b} A{^{IK}}_{c]} \nonu
      \A \A 
      + \ \xi \epsilon^{IJK} 
      (\partial_{[b} A_{0Ic]} 
      + A_{0M [b} A{^M}_{Ic]}). 
      \label{R-dec3}
\ea
The time derivative of the connection appears only 
in (\ref{R-dec1}) and has the form of $\partial_t 
\{ A{_0}{^K}_c - (\xi/2) \epsilon^{IJK} A_{IJc} \}$ 
\cite{Ho}. Thus it is convenient to introduce 
the following variables 
\ba
\A \A {}^- B{^I}_a := A{_0}{^I}_a - {\xi \over 2} \ 
                      \epsilon^{IJK} A_{JKa}, 
                      \label{B+} \\
\A \A {}^+ B{^I}_a := A{_0}{^I}_a + {\xi \over 2} \ 
                      \epsilon^{IJK} A_{JKa}, 
                      \label{B-}
\ea
the inverses of which are given by 
\ba
\A \A A_{0Ia} = {1 \over 2} ({}^- B_{Ia} + {}^+ B_{Ia}), 
                \label{inv1} \\
\A \A A_{IJa} = - {1 \over {2 \xi}} \ \epsilon_{IJK} 
                ({}^- B{^K}_a - {}^+ B{^K}_a). 
                \label{inv2}
\ea
Using (\ref{inv1}) and (\ref{inv2}), 
the covariant derivative $\hat D_a$ of (\ref{Dhat}) 
and the decomposition of the curvature, 
(\ref{R-dec1})-(\ref{R-dec3}), are written 
in terms of ${}^- B{^I}_a$ and ${}^+ B{^I}_a$ as 
\be
\hat D_a = {{1 - i \xi \gamma_5} \over 2} \left\{ 
           \partial_a + \xi^{-1} \left( 
           {{1 - i \xi \gamma_5} \over 2} 
           {}^- B{^I}_a 
           - {{1 + i \xi \gamma_5} \over 2} 
           {}^+ B{^I}_a \right) \gamma_5 S_{0I} 
           \right\}, 
\label{Dhat-B}
\ee
and 
\ba
\hat R{^{0K}}_{tc} = \A \A -{1 \over 2} \partial_t 
                     {}^- B{^K}_c + {1 \over 2} 
                     \partial_c \left( A{_0}{^K}_t 
                     - {\xi \over 2} \epsilon^{IJK} 
                     A_{IJt} \right) \nonu
                     \A \A 
                     + \ \epsilon^{IJK} A_{0It} 
                     \left( {{\xi^2 - 1} \over {4 \xi}} 
                     {}^- B_{Jc} 
                     + {{\xi^2 + 1} \over {4 \xi}} 
                     {}^+ B_{Jc} \right) 
                     - {1 \over 2} A{^{KI}}_t 
                     {}^- B_{Ic}, 
                     \label{Rhat-1} \\
\hat R{^{0K}}_{bc} = \A \A - \partial_{[b} {}^- B{^K}_{c]} 
                     + {{\xi^2 - 3} \over {8 \xi}} 
                     \epsilon^{IJK} 
                     {}^- B_{I [b} 
                     {}^- B_{Jc]} 
                     + {{\xi^2 + 1} \over {4 \xi}} 
                     \epsilon^{IJK} 
                     {}^- B_{I [b} 
                     {}^+ B_{Jc]} \nonu
                     \A \A 
                     + \ {{\xi^2 + 1} \over {8 \xi}} 
                     \epsilon^{IJK} 
                     {}^+ B_{I [b} 
                     {}^+ B_{Jc]}, \\
\hat R{^{JK}}_{bc} = \A \A \epsilon^{IJK} \partial_{[b} 
                     \left( {{\xi^2 - 1} \over {2 \xi}} 
                     {}^- B_{Ic]} 
                     + {{\xi^2 + 1} \over {2 \xi}} 
                     {}^+ B_{Ic]} \right) 
                     + {{3 \xi^2 - 1} \over {4 \xi^2}} 
                     {}^- B{^J}_{[b} 
                     {}^- B{^K}_{c]} \nonu
                     \A \A 
                     + \ {{\xi^2 + 1} \over {2 \xi^2}} 
                     {}^- B{^{[J}}_{[b} 
                     {}^+ B{^{K]}}_{c]} 
                     - {{\xi^2 + 1} \over {4 \xi^2}} 
                     {}^+ B{^J}_{[b} 
                     {}^+ B{^K}_{c]}. 
                     \label{Rhat-3} 
\ea

Then we see from (\ref{Rhat-1}) that ${}^- B{^I}_a$ is 
the dynamical variable, and that the kinetic terms 
in (\ref{3+1dec}) are given by 
\be
- \tilde E_I^a \ \partial_t {}^- B{^I}_a 
- \epsilon^{abc} \ \overline \psi_b \gamma_5 \gamma_c 
{{1 - i \xi \gamma_5} \over 2} \ \partial_t \psi_a, 
\ee
where we have used the identity 
\be
\tilde E_I^a := E E_I^a = {1 \over 2} 
\epsilon^{abc} \epsilon_{IJK} E^J_b E^K_c 
\ee
with $E$ being defined by $E = {\rm det}(E^I_a)$. 
On the other hand, the nondynamical variables 
in (\ref{3+1dec}) are $A_{0It}, A_{IJt}$ 
and ${}^+ B{^I}_a$ in addition to the lapse function 
$N$ and the shift vector $N^a$. 

For $N = 1$ chiral SUGRA with $\xi = \pm i$, 
the variable ${}^+ B{^I}_a$ does not appear 
in (\ref{Dhat-B})-(\ref{Rhat-3}), and 
$(1 - i \xi \gamma_5)/2$ in (\ref{Dhat-B}) becomes 
$(1 \pm \gamma_5)/2$ which generates only the right- 
or left-handed spinor field. 
The dynamical variable ${}^- B{^I}_a$ becomes 
the Ashtekar's one, i.e. ${}^- B{^I}_a \mid_{\xi = \pm i} 
= {}^{{\rm Ash}} A{^I}_a$. 

In case of the non-chiral theory with 
$\xi \not= \pm i$, the constraint corresponding to 
the nondynamical variable ${}^+ B{^I}_a$ appears 
in addition to the constraints obtained 
by varying ${\cal L}$ with respect to 
$A_{0It}, A_{IJt}$; namely, 
\ba
{}^+ P{_I}^a := \A \A {{\delta {\cal L}} \over 
                {\delta {}^+ B{^I}_a}} 
                \ = \ 0, \label{P+} \\
P^{It}:= \A \A {{\delta {\cal L}} \over 
         {\delta A_{0It}}} 
         \ = \ 0, \label{PIt} \\
P^{IJt} := \A \A {{\delta {\cal L}} \over 
           {\delta A_{IJt}}} 
           \ = \ 0. \label{PIJt} 
\ea
The spatial restriction of the Lorentz connection, 
$A_{IJa}$, is determined from only these three 
constraints: In order to show this, we notice 
that $A_{IJa}$ is identically expressed as 
\be
A_{IJa} = A_{IJa}(E, \psi) + {E^K_a \over e} 
(M_{IJK} - 2 M_{K [IJ]} + 2 \delta_{K [I} M_{J]}), 
\label{A-spa}
\ee
where we define 
\be
M{_{ij}}^{\nu} 
:= {\delta {\cal L} \over \delta B{^{ij}}_{\nu}}, 
\label{defM}
\ee
and 
\be
M_{IJK} := E_K^a M_{IJa}, \ \ 
M_I := e^j_{\nu} M{_{Ij}}^{\nu}. 
\ee
In (\ref{A-spa}) $A_{IJa}(E, \psi)$ denotes 
\be
A_{IJa}(E, \psi) := A_{IJa}(E) + \kappa_{IJa} 
\ee
with $A_{IJa}(E)$ being the spatial restriction 
of the Ricci rotation coefficients $A_{ij \mu}(e)$, 
while $\kappa_{IJa}$ being defined as 
\be
\kappa_{IJa} 
:= {i \over 4} (E_I^b E_J^c E^K_a \overline \psi_b 
\gamma_K \psi_c 
+ E_I^b \overline \psi_b \gamma_J \psi_a 
- E_J^b \overline \psi_b \gamma_I \psi_a), 
\ee
which leads to 
\be
\kappa_{I[ba]} := E^J_{[b} \kappa_{IJa]} 
= - {i \over 4} \overline \psi_b \gamma_I \psi_a. 
\ee
If we compare (\ref{defM}) 
with Eqs.(\ref{P+})-(\ref{PIJt}), we can show 
\ba
M_{IJK} = \A \A {\xi \over {\xi^2 + 1}} 
          E^K_a \epsilon^{IJM} {}^+ P{_M}^a, 
          \\
M_I = \A \A - \ {N \over {\xi^2 + 1}} \left( P{_I}^t 
      - {\xi \over 2} \epsilon_{IJK} P^{JKt} \right) 
      + {1 \over {\xi^2 + 1}} N^a E^J_a 
      \left( P{_{IJ}}^t + \xi \epsilon_{IJK} P{_K}^t 
      \right) \nonu
      \A \A + \ {\xi \over {\xi^2 + 1}} E^J_a \epsilon_{IJK} 
      {}^+ P^{Ka}. 
\ea
Thus the constraints give 
\be
A_{IJa} = A_{IJa}(E, \psi). 
\ee
By virtue of (\ref{B+}) and (\ref{B-}), the nondynamical 
variable ${}^+ B{^I}_a$ is now expressed 
by using the dynamical variables as 
\ba
{}^+ B{^I}_a = \A \A {}^- B{^I}_a 
               + \xi \epsilon^{IJK} A_{JKa}(E, \psi) 
               \nonu
             = \A \A {}^- B{^I}_a 
               - 2 \xi \Gamma{^I}_a, 
\label{B+B-}
\ea
where the $SO(3)$ spin connection, $\Gamma{^I}_a$, 
is given by 
\ba
\Gamma{^I}_a := \A \A - \ {1 \over 2} \epsilon^{IJK} 
                A_{JKa}(E, \psi) \nonu
              = \A \A \stackrel{\circ}{\Gamma}{^I}_a(E) 
                - {i \over 8} \ \epsilon^{IJK} 
                \left( E_J^b E_K^c \ \overline \psi_b 
                \gamma_a \psi_c + 2 E_J^b \ 
                \overline \psi_b \gamma_K \psi_a \right) 
\label{Gamma}
\ea
with the spatial Levi-Civita spin connection 
$\stackrel{\circ}{\Gamma}{^I}_a(E) = (-1/2) 
\epsilon^{IJK} E_{Jb} \nabla_a E_K^b$. 

We shall now eliminate the nondynamical variable 
${}^+ B{^I}_a$ from the Lagrangian density ${\cal L}$ 
of (\ref{3+1dec}). The coefficients of 
$A_{0It}$ and $A_{IJt}$, which are denoted 
by $P^{It}$ and $P^{IJt}$, respectively, 
are given by 
\ba
P^{It} 
= \A \A - \ \partial_a \tilde E^{Ia} + \epsilon^{IJK} 
  \left( {{\xi^2 - 1} \over {2 \xi}} {}^- B_{Ja} 
  + {{\xi^2 + 1} \over {2 \xi}} {}^+ B_{Ja} 
  \right) \tilde E_K^a \nonu
  \A \A + \ i \epsilon^{abc} \ \overline \psi_b \gamma_5 
  \gamma_c S{_0}^I {{1 - i \xi \gamma_5} \over 2} \psi_a, 
  \label{delAt} \\
P^{IJt} 
= \A \A {\xi \over 2} \epsilon^{IJK} (\partial_a 
  \tilde E_K^a + \xi^{-1} \epsilon_{KMN} 
  {}^- B{^M}_a \tilde E^{Na} 
  - \xi^{-1} \epsilon^{abc} \ \overline \psi_b \gamma_c 
  S_{0K} {{1 - i \xi \gamma_5} \over 2} \psi_a), 
\ea
and therefore they satisfy 
$P{_I}^t = \xi \ \epsilon_{IJK} P^{IJt}$ 
by virtue of (\ref{B+B-}). 
Then we obtain 
\be
A_{0It} P^{It} + A_{IJt} P^{IJt} 
= \xi \Lambda^I \left( {}^- {\cal D}_a \tilde E_I^a 
- \xi^{-1} \epsilon^{abc} \ \overline \psi_b 
\gamma_c S_{0I} {{1 - i \xi \gamma_5} \over 2} 
\psi_a \right), 
\ee
where $\Lambda^I$ is defined by 
$\Lambda^I := \xi A_{0It} + (1/2) \epsilon_{IJK} 
A{^{JK}}_t$, and the covariant derivative 
is denoted as 
\be
{}^- {\cal D}_a \tilde E_I^a := \partial_a \tilde E_I^a 
+ \xi^{-1} \epsilon_{IJK} {}^- B{^J}_a \tilde E^{Ka}. 
\ee
As for the coefficients of $N$, $N^a$ and $\psi_t$, 
they are obtained by the straightforward calculation. 

Consequently, the Lagrangian density ${\cal L}$ 
is written as 
\ba
{\cal L} = \A \A - \ \tilde E_I^a {}^- \dot B{^I}_a 
           - \epsilon^{abc} \ \overline \psi_b \gamma_5 
           \gamma_c {{1 - i \xi \gamma_5} \over 2} 
           \dot \psi_a \nonu
           \A \A + \ \xi \Lambda^I {\cal G}_I 
           + N^a {\cal V}_a + N {\cal H} + \psi_t {\cal S} 
\label{3+1decF}
\ea
up to boundary terms. In (\ref{3+1decF}), $\Lambda^I$, 
$N^a, N$ and $\psi_t$ are Lagrange multipliers, 
while ${\cal G}_I, {\cal V}_a, {\cal H}$ and ${\cal S}$ 
are the constraints corresponding to these Lagrange 
multipliers, which read as follows: 
\ba
{\cal G}_I := \A \A {}^- {\cal D}_a \tilde E_I^a 
              - \xi^{-1} \epsilon^{abc} \ 
              \overline \psi_b \gamma_c S_{0I} 
              {{1 - i \xi \gamma_5} \over 2} \psi_a 
              \ = \ 0, 
              \label{Gauss} \\
{\cal V}_a := \A \A 2 \tilde E^{Ib} F_{Iab} 
              + \epsilon^{bcd} \ \overline \psi_b 
              \gamma_5 \gamma_a 
              {{1 - i \xi \gamma_5} \over 2} 
              {}^- {\cal D}_c \psi_d \nonu
              \A \A + \ {{i(1 + \xi^2)} \over {2 \xi}} 
              \epsilon^{bcd} \ \overline \psi_b 
              \gamma_0 \psi_c \ K_{[da]} \ = \ 0, 
              \label{vector} \\
{\cal H} := \A \A E^{-1} \epsilon^{IJK} \tilde E_I^a 
            \tilde E_J^b \ \{ \xi F_{Kab} - (1 + \xi^2) 
            R_{Kab} \} \nonu
            \A \A + \ \epsilon^{abc} \ \overline \psi_a 
            \gamma_5 \gamma_0 
            {{1 - i \xi \gamma_5} \over 2} 
            {}^- {\cal D}_b \psi_c \nonu
            \A \A + \ {{i(1 + \xi^2)} \over {4 \xi}} 
            \epsilon^{abc} \ \overline \psi_a 
            \gamma_I \psi_c \ K{^I}_b \ = \ 0, 
            \label{Hamilt} \\
{\cal S} := \A \A - \epsilon^{abc} \gamma_5 \gamma_a 
            {{1 - i \xi \gamma_5} \over 2} 
            {}^- {\cal D}_b \psi_c 
            + {{1 - i \xi \gamma_5} \over 2} 
            {}^- {\cal D}_a 
            (\epsilon^{abc} \gamma_5 \gamma_b 
            \psi_c) \nonu
            \A \A + \ {{i(1 + \xi^2)} \over {2 \xi}} 
            \epsilon^{abc} \gamma_0 \psi_c \ K_{ba} 
            \ = \ 0 \label{SUSY}
\ea
with $K_{ba}$ being given by $K_{ba} = E^I_b K_{Ia} 
:= E^I_b A_{0Ia}$. 
In Eqs.(\ref{vector})-(\ref{SUSY}) 
the covariant derivative ${}^- {\cal D}_a$ 
acts on $\psi_b$ as 
\be
{}^- {\cal D}_a \psi_b := \partial_a \psi_b 
+ \xi^{-1} {}^- B{^I}_a \gamma_5 S_{0I} \ \psi_b, 
\ee
and the curvature tensors are defined by 
\ba
\A \A R_{Iab} := \partial_{[a} \Gamma_{Ib]} 
+ {1 \over 2} \epsilon_{IJK} \Gamma{^J}_{[a} 
\Gamma{^K}_{b]}, \\
\A \A F_{Iab} := \partial_{[a} {}^- B_{Ib]} 
+ {1 \over 2} \xi^{-1} \epsilon_{IJK} {}^- B{^J}_{[a} 
{}^- B{^K}_{b]}. 
\ea
Note that in the vector constraint of (\ref{vector}) 
we have omitted a term proportional 
to the Gauss constraint of (\ref{Gauss}). 

We shall give some comments on the canonical formulation 
described by (\ref{3+1decF}). Firstly, let us give 
the relation of the dynamical variable ${}^- B{^I}_a$ 
to Barbero's or Ashtekar's one. The dynamical variable 
${}^- B{^I}_a$ is written as 
\be
{}^- B{^I}_a = A{_0}{^I}_a 
               - {\xi \over 2} \ \epsilon^{IJK} A_{JKa} 
             = K{^I}_a + \xi \Gamma{^I}_a. 
\label{BKGam}
\ee
This means that the canonical formulation based on 
the Lagrangian density (\ref{3+1decF}) is obtained 
from the usual $N = 1$ SUGRA by the canonical 
transformation (\ref{BKGam}). 
Therefore the ${}^- B{^I}_a$ is related to 
Barbero's dynamical variable, 
${}^{{\rm Bar}} A{^I}_a = \Gamma{^I}_a + \beta K{^I}_a$, 
which now includes the torsion part, by 
\be
{}^- B{^I}_a = \xi \ {}^{{\rm Bar}} A{^I}_a 
\ \ \ {\rm with} \ \ \ \xi = \beta^{-1}. 
\ee
For the chiral case with $\xi = \pm i$, the dynamical 
variable ${}^- B{^I}_a$ becomes Ashtekar's one; namely, 
\be
{}^- B{^I}_a \mid_{\xi = \pm i} 
= K{^I}_a \pm i \Gamma{^I}_a 
= {}^{{\rm Ash}} A{^I}_a. 
\ee

Secondly, we focus our attention on 
those terms proportional to $(1 + \xi^2)$ 
in Eqs.(\ref{vector})-(\ref{SUSY}) 
which violate parity operation. 
As for those of (\ref{vector}) and (\ref{SUSY}), 
the antisymmetric part of $K_{ba}$ is given by 
\be
K_{[ba]} = E^I_{[b} \kappa_{0Ia]} 
         = - {i \over 4} \ \overline \psi_b 
           \gamma_0 \psi_a 
\label{Kasym}
\ee
by virtue of the Gauss constraint (\ref{Gauss}). 
Then we can show that the terms proportional to 
$(1 + \xi^2)$ are canceled by other parity-violating 
terms in (\ref{vector}) and (\ref{SUSY}). 
As for (\ref{Hamilt}), on the other hand, 
if the $K_{ba}$ is given by the `on-shell' expression 
(namely, that in the second-order formalism) as 
a sum of the extrinsic curvature and quadratic 
terms of $\psi_a$, then we can also show that 
four-fermion contact terms in the last term of 
(\ref{Hamilt}) are canceled by other parity-violating 
terms. In case of the chiral theory with 
$\xi = \pm i$, however, those terms proportional to 
$(1 + \xi^2)$ do not appear in the constraints, 
which is one of the advantages in the Ashtekar 
formulation of $N = 1$ SUGRA. 

The final comment is concerned with the dynamical 
variable $\psi_a$ and its conjugate momentum. 
From (\ref{3+1decF}) the conjugate momentum of 
$\psi_a$ is given by 
\be
\pi^a := {{\delta {\cal L}} \over 
          {\delta \dot \psi_a}} 
       = - \epsilon^{abc} \overline 
         \psi_b \gamma_5 \gamma_c 
         {{1 - i \xi \gamma_5} \over 2}. 
\label{MCM}
\ee
However, Eq.(\ref{MCM}) leads to 
the second-class constraint 
\be
\lambda^a := \pi^a + \epsilon^{abc} 
             \overline \psi_b \gamma_5 \gamma_c 
             {{1 - i \xi \gamma_5} \over 2} 
       \ \ = \ \ 0 
\ee
unless $\xi = \pm i$. 
Therefore, in case of the non-chiral theory, 
we must compute the Dirac brackets among the basic field 
variables in order to eliminate $\lambda^a$ 
as in the usual $N = 1$ SUGRA \cite{Pila,DEa}. 
Moreover, if we try to make the Dirac bracket 
of $({}^- B{^I}_a, \psi_b)$, 
or of $({}^- B{^I}_a, {}^- B{^J}_b)$ vanish, 
we will have to change the form of ${}^- B{^I}_a$. 

Recently, Thiemann has proposed the Lorentzian 
Hamiltonian constraint of spin-1/2 fields, in which 
the fermions is treated as half-densities \cite{TT2}. 
For spin-3/2 fields, a similar approach is to take 
the weighted tetrad component, 
$\phi_I := E^{1/2} E_I^a \ \psi_a$, as a basic 
field variable \cite{DKS}. As preliminaly, 
the canonical transformation 
\be
{}^- \stackrel{\circ}{B}{^I}_a := K{^I}_a 
+ \xi \stackrel{\circ}{\Gamma}{^I}_a(E) 
\ee
has been considered \cite{SKpri}, 
where $\stackrel{\circ}{\Gamma}{^I}_a(E)$ is 
the spatial Levi-Civita spin connection 
defined in (\ref{Gamma}). Contratry to the case of 
spin-1/2 fields, however, it has been found that 
the form of ${}^- \stackrel{\circ}{B}{^I}_a$ 
must be changed in order to make the Dirac bracket 
of $({}^- \stackrel{\circ}{B}{^I}_a, \psi_b)$, 
or of $({}^- \stackrel{\circ}{B}{^I}_a, 
{}^- \stackrel{\circ}{B}{^J}_b)$ vanish. 

To summarize, in this paper we have generalized 
the Lagrangian of $N = 1$ SUGRA by using an arbitrary 
parameter $\xi$ as the extension of the pure-gravity case 
\cite{Ho}. This generalized Lagrangian gives 
the canonical formulation with the real Ashtekar variable 
after the 3+1 decomposition of spacetime. 
The constraints in this formulation are also derived 
from those of the usual $N = 1$ SUGRA 
by performing Barbero's type canonical transformation 
with an arbitrary parameter $\beta \ (= \xi^{-1})$. 
In particular, for $\xi = \pm i$, the formulation 
of this paper is equivalent with the chiral one 
\cite{Jac}. The detailed analysis for canonical 
quantization of $N = 1$ SUGRA with the real 
Ashtekar variable needs future investigation.

{\large{\bf{Acknowledgments}}} 

I am grateful to Professor T. Shirafuji 
for discussions and reading the manuscript. 
Also, I would like to thank the members 
of Physics Department at Saitama University 
for discussions and encouragements.




\newpage
\begin{thebibliography}{100}

\bibitem{AA} 
A. Ashtekar, 
{\it Phys. Rev. Lett.} {\bf 57}, 2244 (1986); 
{\it Phys. Rev.} D {\bf 36}, 1587 (1987). 

\bibitem{AA2} 
A. Ashtekar, 
{\it Lectures on Non-perturbative Canonical Gravity} 
(World Scientific, Singapore 1991). 

\bibitem{Bar} 
J. F. Barbero G., 
{\it Phys. Rev.} D {\bf 51}, 5507 (1995). 

\bibitem{ALMMT} 
A. Ashtekar, J. Lewandowski, D. Marolf, 
J. Mourao and T. Thiemann, 
{\it J. Math. Phys.} {\bf 36}, 6456 (1995). 

\bibitem{TT1} 
T. Thiemann, 
{\it Phys. Lett.} {\bf 380B}, 257 (1996); 
{\it Class. Quantum Grav.} {\bf 15}, 839 (1998). 

\bibitem{Ho} 
S. Holst, 
{\it Phys. Rev.} D {\bf 53}, 5966 (1996). 

\bibitem{Imm} 
G. Immirzi, 
{\it Nucl. Phys. Proc. Suppl.} {\bf 57}, 65 (1997); 
{\it Class. Quantum Grav.} {\bf 14}, L177 (1997). 

\bibitem{JS} 
T. Jacobson and L. Smolin, 
{\it Phys. Lett.} {\bf 196B}, 39 (1987); 
{\it Class. Quantum Grav.} {\bf 5}, 583 (1988). 

\bibitem{TSX} 
M. Tsuda, T. Shirafuji and H. Xie, 
{\it Class. Quantum Grav.} {\bf 12}, 3067 (1995). 

\bibitem{Mi1} 
E. W. Mielke, A. Mac\'ias and H. A. Morales-T\'ecotl, 
{\it Phys. Lett.} {\bf 215A}, 14 (1996). 

\bibitem{Mi2} 
E. W. Mielke and A. Mac\'ias, 
Report No. gr-qc/9902077. 

\bibitem{Jac} 
T. Jacobson, 
{\it Class. Quantum Grav.} {\bf 5}, 923 (1988). 

\bibitem{TS} 
M. Tsuda and T. Shirafuji, 
{\it Phys. Rev.} D {\bf 54}, 2960 (1996). 

\bibitem{UGOP} 
D. Armand-Ugon, R. Gambini, O. Obreg\'on and J. Pullin, 
{\it Nucl. Phys.} B {\bf 460}, 615 (1996). 

\bibitem{Pila} 
M. Pilati, 
{\it Nucl. Phys.} B {\bf 132}, 138 (1978). 

\bibitem{DEa} 
P. D. D'Eath, 
{\it Phys. Rev.} D {\bf 29}, 2199 (1984). 

\bibitem{TT2} 
T. Thiemann, 
{\it Class. Quantum Grav.} {\bf 15}, 1281 (1998). 

\bibitem{DKS} 
S. Deser, J. H. Kay and K. S. Stelle, 
{\it Phys. Rev.} D {\bf 16}, 2448 (1977). 

\bibitem{SKpri} 
M. Sawaguchi and C. Kim (private communication). 









\end{thebibliography}
\end{document}